\documentclass[12pt]{iopart}

\usepackage{amsfonts}
\usepackage{amssymb}
\usepackage{lmodern,dsfont}
\usepackage{graphicx}
\usepackage[usenames,dvipsnames]{xcolor}
\usepackage{bm}
\usepackage{blkarray}
\usepackage{blindtext}
\usepackage[english=american]{csquotes}
\usepackage{hyperref}

\newcommand{\Av}[1]{\left\langle #1 \right\rangle}
\newcommand{\av}[1]{\langle #1 \rangle}

\newcommand{\n}{\nonumber}
\newcommand{\nn}{\nonumber \\}

\newcommand{\eqref}{\eref}
\newcommand{\text}{\mathrm}

\begin{document}
\title[Multidimensional thermodynamic uncertainty relations]{Multidimensional thermodynamic uncertainty relations}
\author{Andreas Dechant}
\address{WPI-Advanced Institute of Materials Research (WPI-AIMR), Tohoku University, Sendai 980-8577, Japan}
\ead{andreas.dechant@outlook.com}

\begin{abstract}
We extend a class of recently derived thermodynamic uncertainty relations to vector-valued observables.
In contrast to the scalar-valued observables examined previously, this multidimensional thermodynamic uncertainty relation provides a natural way to study currents in high-dimensional systems and to obtain relations between different observables.
Our proof is based on the generalized Cr{\'a}mer-Rao inequality, which we interpret as a relation between physical observables and the Fisher information.
This allows us to develop high-dimensional versions of both the original, steady state uncertainty relation and the more recently obtained generalized uncertainty relation for time-periodic systems.
We apply the multidimensional uncertainty relation to obtain a new constraint on the performance of steady-state heat engines, which is tighter than previous bounds and reveals the role of heat-work correlations.
As a second application, we show that the uncertainty relation is connected to a bound on the differential mobility.
As a result of this connection, we find that a necessary condition for equality in the uncertainty relation is that the system obeys the equilibrium fluctuation-dissipation relation.
\end{abstract}

\maketitle

\newenvironment{align}{\begin{eqnarray}}{%
    \end{eqnarray}\ignorespacesafterend
}

A remarkable property of stochastic transport in steady-state systems is the so-called thermodynamic uncertainty relation (TUR), which was first conjectured by Barato et al. \cite{Bar15} and subsequently proven by Gingrich et al. \cite{Gin16}.
This relation states that the square of the average current is bounded from above by the variance of the current times the entropy production.
It thus provides a universal relation between a current, its fluctuations and the thermodynamic cost of driving the current.
This relation was later generalized from the long-time limit to steady state systems at finite time \cite{Pie17,Hor17,Dec17} and to time-periodic systems \cite{Bar18,Koy18}.
In all these formulations, the stochastic current is a scalar quantity.
While many toy models involve one-dimensional settings, where considering scalar observables is sufficient, realistic physical systems are placed in three-dimensional space, where both the direction and the magnitude of a current may be important.
Further, one may be interested in the behavior of several currents, which are generally correlated.
Thus, it is desirable to obtain a bound in the spirit of the TUR, which explicitly takes into account the dimensionality of the system.

In this work, we extend the family of TURs to vector-valued observables.
In the process, we will re-derive the TUR from a central result of information theory, the Cr{\'a}mer-Rao bound \cite{Rao45,Cra16}.
This derivation shows that the TUR can be understood as a consequence of the information-geometric properties of the space of path probability densities \cite{Has18}.
The resulting multidimensioal thermodynamic uncertainty relation (MTUR) encompasses both the finite-time result for steady states \cite{Pie17,Hor17,Dec17} and the more recently obtained result for time-periodic systems \cite{Bar18,Koy18}.
The derivation from the Cr{\'a}mer-Rao bound not only allows for a straightforward extension to multidimensional systems with vector-valued observables.
It further shows that the result obtained in Ref.~\cite{Koy18} is not restricted to time-periodic driving, but also holds for arbitrary time-dependent systems under the assumption of certain regularity conditions.
Throughout this paper, we will focus on Langevin dynamics, however, we stress that all the results also hold for Markov jump dynamics, with the specific expressions given in the Appendix.

\section{Fisher information and the Cr{\'a}mer-Rao bound} \label{sec-cramer-rao}
The mathematical foundation for our results is the generalized Cr{\'a}mer-Rao bound.
Suppose that we have a probability density $P(\omega,\bm{\theta}) = P(\Omega = \omega \vert \bm{\theta})$ for some general random variable $\Omega$ and depending on a set of $M$ parameters $\bm{\theta} = \lbrace \theta_1, \ldots, \theta_M \rbrace$.
Note that $\Omega$ may present the instantaneous value of some stochastic process $\bm{x}(t)$, but may equally well be the path of the stochastic process $\lbrace \bm{x}(t) \rbrace_{t \in [0,\mathcal{T}]}$ during the time interval $[0,\mathcal{T}]$.
We further consider a set of $K$ observables $\bm{r}(\Omega) = \lbrace r_1(\Omega), \ldots, r_K(\Omega) \rbrace$ with average $\av{\bm{r}}_\theta = \int d\omega \ \bm{r}(\omega) P(\omega,\bm{\theta})$. 
Depending on the random variable $\Omega$, the integral $d\omega$ can represent a sum over discrete states, an integral over a set of continuous variables or a path integral.
The generalized Cr{\'a}mer-Rao bound is written as an operator inequality \cite{Kay93}
\begin{eqnarray}
\bm{J}_r(\bm{\theta})^T \bm{\Xi}_r(\bm{\theta})^{-1} \bm{J}_r(\bm{\theta}) \leq \bm{I}(\bm{\theta}), \label{cramer-rao}
\end{eqnarray}
where $T$ denotes transposition and we defined the $K \times M$ Jacobian $\bm{J}_r(\bm{\theta})$ of $\av{\bm{r}}_\theta$ with respect to $\bm{\theta}$,
\begin{eqnarray}
\big(\bm{J}_r(\bm{\theta})\big)_{i j} = \partial_{\theta_j} \av{r_i}_\theta, \label{jacobian}
\end{eqnarray}
the positive definite $K \times K$ covariance matrix $\bm{\Xi}_r(\bm{\theta})$,
\begin{eqnarray}
\big(\bm{\Xi}_r(\bm{\theta})\big)_{i j} = \av{r_i r_j}_\theta - \av{r_i}_\theta \av{r_j}_{\theta}, \label{covariance}
\end{eqnarray}
and the positive semidefinite $M \times M$ Fisher information matrix
\begin{eqnarray}
\big(\bm{I}(\bm{\theta})\big)_{i j} = \int d\omega \ \frac{\partial_{\theta_i} P(\omega,\bm{\theta}) \partial_{\theta_j} P(\omega,\bm{\theta})}{P(\omega,\bm{\theta})}  \label{fisher-info}.
\end{eqnarray}
The operator inequality \eref{cramer-rao} is interpreted as $\bm{I} - \bm{J}_r^T \bm{\Xi}_r^{-1} \bm{J}_r$ being a positive semidefinite matrix, i.~e.~for some arbitrary vector $\bm{v} \in \mathbb{R}^M$ we have $\bm{v}^T (\bm{I} - \bm{J}_r^T \bm{\Xi}_r^{-1} \bm{J}_r) \bm{v} \geq 0$.
The physical interpretation of the generalized Cr{\'a}mer-Rao bound is most conveniently made clear by focusing on the case of a single observable $r$ and parameter $\theta_i$, in which case Eq.~\eref{cramer-rao} simplifies to the inequality
\begin{eqnarray}
\frac{\big(\partial_{\theta_i} \av{r}_\theta\big)^2}{\av{\Delta r^2}_\theta} \leq \big(\bm{I}(\bm{\theta}) \big)_{ii} .
\end{eqnarray}
The left-hand side is the change in the average of the observable $r$ due to a change in the parameter $\theta_i$, relative to the fluctuations of $r$.
The left-hand thus side tells us how much information on the parameter $\theta_i$ the observable $r$ contains:
If the average of $r$ is almost independent of $\theta_i$ (or the fluctuations of $r$ are very large), then a measurement of $r$ will not allow us to make any statement about the value of $\theta_i$.
On the other hand, if $\av{r}_\theta$ changes substantially by varying $\theta_i$, then the measurement of $r$ can potentially provide us with a good estimate of $\theta_i$.
The Cr{\'a}mer-Rao bound states that this information on $\theta_i$ contained in any observable is always less then the corresponding Fisher information $\bm{I}_{ii}$.
In other words, the probability density itself contains the maximum amount of information; measuring any observable can only yield less information.

\section{Path Fisher information for Langevin dynamics} \label{sec-fisher-path}
To make the connection between these information-theoretic ideas and a concrete physical situation, consider the $N$-dimensional diffusion process $\bm{x}(t) = \lbrace x_1(t), \ldots, x_N(t) \rbrace$ described by the It{\=o}-Langevin equation \cite{Ris86},
\begin{eqnarray}
\dot{\bm{x}}(t) = \bm{a}(\bm{x}(t),t,\bm{\theta}) + \sqrt{2 \bm{B}(\bm{x}(t),t)} \cdot \bm{\xi}(t), \label{langevin}
\end{eqnarray}
with mutually independent Gaussian white noises $\xi_i(t)$.
Equivalently, we have the Fokker-Planck equation for the probability density $P(\bm{x},t,\bm{\theta})$ and current $\bm{j}(\bm{x},t,\bm{\theta})$ \cite{Ris86},
\begin{eqnarray}
\partial_t P(\bm{x},t,\bm{\theta}) = - \bm{\nabla} \bm{j}(\bm{x},t,\bm{\theta}) \label{fpe-org} \\
 \bm{j}(\bm{x},t,\bm{\theta}) = \Big( \bm{a}(\bm{x},t,\bm{\theta}) - \bm{\nabla} \bm{B}(\bm{x},t) \Big) P(\bm{x},t,\bm{\theta}) \n
\end{eqnarray}
with $K$-dimensional drift vector (or generalized forces) $\bm{a}(\bm{x},t,\bm{\theta})$ and symmetric, positive definite $N \times N$ diffusion matrix $\bm{B}(\bm{x},t)$.
We assume that the generalized forces $\bm{a}(\bm{x},t,\bm{\theta})$ depend on a set of control parameters $\bm{\theta}$ as
\begin{eqnarray}
\bm{a}(\bm{x},t,\bm{\theta}) &= \bm{a}_0(\bm{x},t) + \sum_{i=1}^M \theta_i \bm{a}_i(\bm{x},t) . 
\end{eqnarray}
Physically, we may take $\bm{a}_0$ to define a reference system and the $\bm{a}_i$ to be perturbations to this reference system.
For small values of the parameters $\bm{\theta}$ the Fisher information matrix then describes the linear response behavior of the system.
We define the random variable $\omega$ as the path $\lbrace \bm{x}(t) \rbrace_{t \in [0,\mathcal{T}]}$ of the diffusion process corresponding to the above Langevin dynamics.
The probability density of the path is given by the Onsager-Machlup functional \cite{Ris86}
\begin{eqnarray}
&\mathbb{P}\big(\lbrace \bm{x}(t) \rbrace_{t \in [0,\mathcal{T}]}\big) \propto \exp\big[-\mathcal{S}[\bm{x}(t),\bm{\theta}] \big] P_0(\bm{x}(0),\bm{\theta}) \\
&\quad \mathrm{with} \qquad \mathcal{S}[\bm{x}(t),\bm{\theta}] = \frac{1}{4} \int_0^\mathcal{T} dt \ \big(\dot{\bm{x}}(t) - \bm{a}(t) \big)^T \bm{B}(t)^{-1} \big(\dot{\bm{x}}(t) - \bm{a}(t) \big) \n ,
\end{eqnarray}
and a prefactor that depends only on the diffusion matrix $\bm{B}$.
Note that here and in the following we use the short-hand notation $f(t) = f(\bm{x}(t),t)$ to denote functions evaluated along the trajectory.
For now, we ignore a possible dependence of the initial state $P_0(\bm{x},\bm{\theta})$ on the parameters $\bm{\theta}$.
Then, the derivative of the path probability with respect to $\theta_i$ is given by
\begin{eqnarray}
&\frac{\partial_{\theta_i} \mathbb{P}\big(\lbrace \bm{x}(t) \rbrace_{t \in [0,\mathcal{T}]}\big)}{ \mathbb{P}\big(\lbrace \bm{x}(t) \rbrace_{t \in [0,\mathcal{T}]}\big)} = \frac{1}{2} \int_0^\mathcal{T} dt \ \bm{a}_i(t)^T \bm{B}(t)^{-1} \big(\dot{\bm{x}}(t) - \bm{a}(t) \big) \label{path-derivative}  ,
\end{eqnarray}
where we used the symmetry of $\bm{B}$.
The Fisher information matrix of the path probability density is then given by the following path integral
\begin{eqnarray}
\big(\mathbb{I}(\bm{\theta})\big)_{i j} = \frac{1}{4} \int \mathcal{D}\hspace{-.02cm}\bm{x}(t) &\bigg(\int_0^\mathcal{T} dt \int_0^\mathcal{T} ds \ \bm{a}_i(t)^T \bm{B}(t)^{-1} \big(\dot{\bm{x}}(t) - \bm{a}(t) \big) \\
& \quad \times \bm{a}_j(s)^T \bm{B}(s)^{-1} \big(\dot{\bm{x}}(s) - \bm{a}(s) \big) \bigg) \mathbb{P}\big(\lbrace \bm{x}(t) \rbrace_{t \in [0,\mathcal{T}]}\big) \n .
\end{eqnarray}
Along any path of the diffusion process, we have $\dot{\bm{x}}(t) - \bm{a}(t) = \sqrt{2 \bm{B}(t)} \cdot \bm{\xi}(t)$, and thus
\begin{eqnarray}
\big(\mathbb{I}(\bm{\theta})\big)_{i j} = \frac{1}{2} \int \mathcal{D}\hspace{-.02cm}\bm{x}(t) &\bigg(\int_0^\mathcal{T} dt \int_0^\mathcal{T} ds \ \bm{a}_i(t)^T \sqrt{\bm{B}(t)^{-1}} \bm{\xi}(t) \\
& \quad \times \bm{\xi}(s)^T \sqrt{\bm{B}(s)^{-1}} \bm{a}_j(s) \bigg) \mathbb{P}\big(\lbrace \bm{x}(t) \rbrace_{t \in [0,\mathcal{T}]}\big) \n .
\end{eqnarray}
Since the noises are white and uncorrelated, we can write
\begin{eqnarray}
\bm{\xi}(t) \bm{\xi}(s)^T = \bm{1} \delta(t-s),
\end{eqnarray}
where $\bm{1}$ is the $N \times N$ identity matrix.
Thus, the integral over $s$ becomes trivial,
\begin{eqnarray}
\big(\mathbb{I}&(\bm{\theta})\big)_{i j} = \frac{1}{2} \int \mathcal{D}\hspace{-.02cm}\bm{x}(t)\bigg( \int_0^\mathcal{T} dt \ \bm{a}_i(t)^T \bm{B}(t)^{-1} \bm{a}_j(t) \bigg) \mathbb{P}\big(\lbrace \bm{x}(t) \rbrace_{t \in [0,\mathcal{T}]}\big) .
\end{eqnarray}
Since the quantity over which the path integral is performed only depends on the single time $t$, we can replace the path probability density with the one-time probability density and obtain,
\begin{eqnarray}
\big(\mathbb{I}&(\bm{\theta})\big)_{i j} =\frac{1}{2} \int_0^\mathcal{T} dt \ \Av{\bm{a}_i^T \bm{B}^{-1} \bm{a}_j}_{t,\theta} + \big(\bm{I}_0(\bm{\theta})\big)_{i j} \label{fisher-path},
\end{eqnarray}
where the average $\av{\ldots}_{t,\theta}$ is taken with respect to the solution of Eq.~\eref{fpe-org} with parameter values $\bm{\theta}$ and $\bm{I}_0(\bm{\theta})$ is the Fisher information matrix of the initial state.
For the case of a single parameter, this expression was recently obtained in Ref.~\cite{Has18}.
The Fisher information matrix of the path probability density is thus explicitly expressed in terms of the perturbing generalized forces $\bm{a}_i$.
By the Cr{\'a}mer-Rao bound \eref{cramer-rao}, this quantity bounds the response of \emph{any} path-dependent observable $\bm{r}[\bm{x}(t)]$ to the perturbations $\bm{a}_i$.
We can write the Cr{\'a}mer-Rao bound in a more intuitive way by defining
\begin{eqnarray}
d \av{\bm{r}}_\theta = \av{\bm{r}}_{\theta + d\theta} - \av{\bm{r}}_\theta = \bm{J}_r(\bm{\theta}) d\bm{\theta},
\end{eqnarray}
which results in
\begin{eqnarray}
d \av{\bm{r}}_\theta^T \bm{\Xi}_r(\bm{\theta})^{-1} d \av{\bm{r}}_\theta \leq d\bm{\theta}^T \bm{I}(\bm{\theta}) d\bm{\theta} .
\end{eqnarray}
The left-hand side is the response of the observable $\bm{r}$ relative to its fluctuations.
The right-hand side can be related to another information-theoretic quantity, the Kullback-Leibler divergence between two probability densities $P$ and $Q$
\begin{eqnarray}
D_\mathrm{KL}(P \vert Q) = \int d\omega \ P \ln\bigg(\frac{P}{Q}\bigg) .
\end{eqnarray}
The Kullback-Leibler divergence is positive and vanishes only at $P = Q$, its global minimum.
The curvature around the minimum is given by the Fisher information matrix,
\begin{eqnarray}
D_\mathrm{KL}(P(\bm{\theta} + d\bm{\theta}) \vert P(\bm{\theta})) = \frac{1}{2} d\bm{\theta}^T \bm{I}(\bm{\theta}) d\bm{\theta} + O(d\theta^3) 
\end{eqnarray}
and we thus arrive at
\begin{eqnarray}
d \av{\bm{r}}_\theta^T \bm{\Xi}_r(\bm{\theta})^{-1} d \av{\bm{r}}_\theta \leq 2 D_\mathrm{KL}(P(\bm{\theta} + d\bm{\theta}) \vert P(\bm{\theta})) \label{FRI} .
\end{eqnarray}
This is the extension of the fluctuation-response inequality (FRI) derived in Ref.~\cite{Dec18B} to vector-valued observables and more than one perturbation.
In the multidimensional case, the variance of the observable $\bm{r}$ is replaced by its covariance matrix.
This relation can be understood as the connection between the macroscopic and microscopic response of the system to a change of the parameters $\bm{\theta}$.
The left-hand side is the response of some observable (i.~e.~a macroscopic, ensemble-averaged quantity) relative to its fluctuations.
By contrast, the right hand side quantifies the change in the path probability (i.~e.~the microscopic dynamics) as a result of the perturbation.
The FRI thus states that the response of any macroscopic observable is bounded by the change in the microscopic dynamics.

\section{Multidimensional thermodynamic uncertainty relations} \label{sec-uncertainty}
In Ref.~\cite{Dec18B} the steady state thermodynamic uncertainty relation (TUR) was derived from the FRI \eref{FRI} by using a special choice for the perturbation.
Here, we extend this derivation to the multidimensional case and explicitly time-dependent dynamics.
The type of observables described by the TUR are time-integrated currents, defined by
\begin{eqnarray}
\bm{r}[\bm{x}(t)] = \int_0^\mathcal{T} dt \ \bm{Z}(\bm{x}(t),t) \circ \dot{\bm{x}}(t) \label{current},
\end{eqnarray}
where $\bm{Z}$ is an arbitrary $K \times N$ matrix-valued function and $\circ$ denotes the Stratonovich-product.
The average of such a time-integrated current is given in terms of the probability current $\bm{j}$,
\begin{eqnarray}
\av{\bm{r}}_\theta &= \int_0^\mathcal{T} dt \int d\bm{x} \ \bm{Z}(\bm{x},t) \bm{j}(\bm{x},t,\bm{\theta}) .
\end{eqnarray}
We now consider a single perturbation given by
\begin{align}
\bm{a}_1(\bm{x},t) = \frac{\big(\bm{a}_0(\bm{x},t) - \bm{\nabla} \bm{B}(\bm{x},t) \big) \mathcal{P}(\bm{x})}{P(\bm{x},t,0)},
\end{align}
with an arbitrary, time-independent probability density $\mathcal{P}(\bm{x}) > 0$, $\int d\bm{x} \ \mathcal{P}(\bm{x}) = 1$.
For this choice, the Fokker-Planck equation \eqref{fpe-org} reads
\begin{align}
\partial_t P(\bm{x},t,\theta) = - \bm{\nabla} \Bigg( \bm{a}_0(\bm{x},t) + \theta &\frac{\big(\bm{a}_0(\bm{x},t) - \bm{\nabla} \bm{B}(\bm{x},t) \big) \mathcal{P}(\bm{x})}{P(\bm{x},t,0)} \\
 &- \bm{\nabla} \bm{B}(\bm{x},t) \Bigg) P(\bm{x},t,\theta) . \n
\end{align}
Formally, this equation is solved by
\begin{align}
P(\bm{x},t,\theta) = P(\bm{x},t,0) + \theta \Big( P(\bm{x},t,0) - \mathcal{P}(\bm{x}) \Big) .
\end{align}
However, while this solution is normalized, it is not positive for arbitrary $\mathcal{P}$ and $\theta$.
The reason is that, for an arbitrary choice of $\mathcal{P}$, the drift coefficient $\bm{a}_1$ can become very large at points where the probability density in the unperturbed system is small and thus the dynamics is no longer well-defined.
Demanding that the solution should be a proper positive probability density restricts the possible choices of $\mathcal{P}$,
\begin{align}
\mathcal{P}(\bm{x}) < \frac{1+\theta}{\theta} P(\bm{x},t,0) .
\end{align}
This typically can be satisfied by a normalized probability density only for sufficiently small $\theta \ll 1$.
Supposing we have such a $\mathcal{P}$, the probability current is given by
\begin{align}
\bm{j}(\bm{x},t,\theta) = (1+\theta) \bm{j}(\bm{x},t,0),
\end{align}
i.~e.~the additional drift vector leads to a rescaling of the probability currents.
The same is then obviously true for the average of the time-integrated current Eq.~\eqref{current},
\begin{align}
\partial_\theta \av{\bm{r}}_\theta = \av{\bm{r}}_0 .
\end{align}
From Eqs.~\eqref{cramer-rao} and \eqref{fisher-path}, we then immediately have the multidimensional version of the generalized thermodynamic uncertainty relation (GTUR),
\begin{align}
\av{\bm{r}}^T \bm{\Xi}_r^{-1} \av{\bm{r}} \leq \frac{1}{2} \Sigma,
\end{align}
where the quantity $\Sigma$ has a structure similar to the entropy production,
\begin{align}
\Sigma = \int_0^\mathcal{T} dt \int d\bm{x} \ \frac{\bm{\nu}^T(\bm{x},t) \bm{B}^{-1}(\bm{x},t) \bm{\nu}(\bm{x},t)}{P(\bm{x},t)} \\
\text{with} \quad \bm{\nu}(\bm{x},t) = \Big(\bm{a}_0(\bm{x},t) - \bm{\nabla} \bm{B}(\bm{x},t) \Big) \mathcal{P}(\bm{x}) \n .
\end{align}
For a scalar current and periodic driving, this GTUR has been derived for a jump process in Ref.~\cite{Koy18}.
The above derivation shows that the GTUR has a straightforward extension to vector-valued currents by replacing the variance of the current with the covariance matrix.
Further, the result holds not only for periodic but for arbitrary time-dependent systems.
We remark that for a time-periodic system with $P(\bm{x},\mathcal{T}) = P(\bm{x},0)$ and a current without explicit time-dependence $\bm{Z}(\bm{x},t) \equiv \bm{Z}(\bm{x})$, we obtain the bound derived in Ref.~\cite{Bar18} for the choice
\begin{align}
\bm{a}_1(\bm{x},t) = \frac{\int_0^\mathcal{T} dt \ \bm{j}(\bm{x},t)}{\mathcal{T} P(\bm{x},t)},
\end{align}
yielding the inequality
\begin{align}
\av{\bm{r}}^T \bm{\Xi}_r^{-1} \av{\bm{r}} \leq \frac{1}{2} \overline{\Sigma}
\end{align}
with the quantity $\overline{\Sigma}$ defined in terms of the time-averaged probability current
\begin{align}
\overline{\Sigma} = \int_0^\mathcal{T} dt \int d\bm{x} \ \frac{\bar{\bm{j}}^T(\bm{x}) \bm{B}^{-1}(\bm{x},t) \bar{\bm{j}}(\bm{x})}{P(\bm{x},t)} \\
 \text{with} \quad \bar{\bm{j}}(\bm{x}) = \frac{1}{\mathcal{T}} \int_0^\mathcal{T} dt \ \bm{j}(\bm{x},t) . \n
\end{align}

For steady-state systems, we can choose $\mathcal{P}(\bm{x}) = P^\text{st}(\bm{x})$.
For this choice, we have $\bm{\nu}(\bm{x}) = \bm{j}^\text{st}(\bm{x})$ and the quantity $\Sigma$ is precisely the entropy production $\Delta S$ during the time interval $[0,\mathcal{T}]$, thus yielding the steady-state multidimensional thermodynamic uncertainty relation (MTUR)
\begin{eqnarray}
\av{\bm{r}}^T &\bm{\Xi}_r^{-1} \av{\bm{r}} \leq \frac{1}{2} \Delta S \label{uncertainty-steady} .
\end{eqnarray}
For scalar currents, this inequality has been extensively discussed in the literature \cite{Bar15,Gin16,Pie17,Hor17,Dec17}.
The present generalization to vector-valued currents has several advantages.
First, it allows discussing currents in systems with more than one spatial dimension in a natural way.
Second, the bound provided by Eq.~\eqref{uncertainty-steady} is tighter than the bound on any scalar current formed by a linear combination of the individual currents, see below.
Finally, the MTUR explicitly takes into account correlations, providing additional insight into the relation  between different currents.
If the observables $r_i$ are independent of each other, then their covariance matrix is diagonal, $\av{\Delta r_i \Delta r_j} = \delta_{i j} \av{\Delta r_i^2}$, and we obtain
\begin{eqnarray}
\sum_i \frac{\av{r_i}^2}{\av{\Delta r_i^2}} \leq \frac{1}{2} \Delta S .
\end{eqnarray}
Since all the terms on the left-hand side are positive, it is obvious that each of the observables obeys the uncertainty relation on its own.
However, the bound on the sum is obviously tighter.
For two observables $r_1$ and $r_2$, we can write the bound explicitly,
\begin{eqnarray}
\av{\Delta r_2^2} \av{r_1}^2 - 2 \av{\Delta r_1 \Delta r_2} &\av{r_1} \av{r_2} +  \av{\Delta r_1^2} \av{r_2}^2 \nn
 &\leq \frac{1}{2} \Big(\av{\Delta r_1^2} \av{\Delta r_2^2} - \av{\Delta r_1 \Delta r_2}^2 \Big) \Delta S \label{uncertainty-2dim} .
\end{eqnarray}
This bound involves the variances of the individual currents as well as their correlation.
We will discuss some consequence of the explicit dependence on the correlations in the next section.
We remark that we may also obtain a joint bound on $\av{r_1}$ and $\av{r_2}$ by considering the scalar observable $\rho(\bm{x}) = \cos(\varphi) r_1(\bm{x}) + \sin(\varphi) r_2(\bm{x})$, i.~e.~the projection of $\bm{r}$ onto an arbitrary unit vector. 
The corresponding scalar uncertainty relation $\av{\rho}^2/\av{\Delta \rho^2} \leq \Delta S/2$ then yields a bound that is generally less tight than Eq.~\eqref{uncertainty-2dim} and tends to Eq.~\eqref{uncertainty-2dim} upon maximization with respect to $\varphi$.
Thus the bound on vector-valued observables Eq.~\eqref{uncertainty-steady} is always tighter than the bound on any scalar formed by a linear combination of the entries of $\bm{r}$.

\section{Consequences of the multidimensional TUR}

\subsection{Tradeoff relations}
One of the consequences of the TUR is a tradeoff relation between power and efficiency for steady-state heat engines \cite{Pie16,Dec17,Koy18,Pie18,Vro18}.
For an engine operating between two heat baths at temperatures $T_\text{c}$ and $T_\text{h} > T_\text{c}$, the steady-state entropy production rate $\sigma^\text{st} = \Delta S/\mathcal{T}$ can be written as
\begin{align}
\sigma^\text{st} = - \frac{1}{T_\text{c}} \av{\dot{q}_\text{c}} + \frac{1}{T_\text{h}} \av{\dot{q}_\text{h}},
\end{align}
where $q_\text{c}$ and $q_\text{h}$ are the amounts of heat dissipated into the cold, respectively absorbed from the hot heat bath, both of which are time-integrated currents of the type Eq.~\eqref{current}.
In terms of the power output of the engine $\av{\dot{w}} = \av{\dot{q}_\text{h}} - \av{\dot{q}_\text{c}}$ and the Carnot efficiency $\eta_\text{C} = 1 - T_\text{c}/T_\text{h}$, this can be written as
\begin{align}
\sigma^\text{st} = -\frac{1}{T_\text{c}} \big( \av{\dot{w}} - \eta_\text{c} \av{\dot{q_\text{h}}} \big) = \frac{\av{\dot{q}_\text{h}}}{T_\text{c}} \big( \eta_\text{C} - \eta \big),
\end{align}
where we introduced the efficiency $\eta = \av{\dot{w}}/\av{\dot{q}_\text{h}}$.
Obviously, the condition that $\sigma^\text{st} \geq 0$ implies $\eta \leq \eta_\text{C}$ if the engine is supposed to perform work at a positive rate.
From the one-dimensional TUR for the heat current $\dot{q}_\text{h}$ and work current $\dot{w}$ respectively, we have
\begin{align}
\av{\dot{q}_\text{h}}^2 \leq D_{q q} \sigma^\text{st} = \frac{D_{qq}}{T_\text{c}} \av{\dot{q}_\text{h}} \big(\eta_\text{C} - \eta \big) \\
\av{\dot{w}}^2 \leq D_{w w} \sigma^\text{st} = \frac{D_{ww}}{T_\text{c}} \frac{\av{\dot{w}}}{\eta} \big(\eta_\text{C} - \eta \big) \n ,
\end{align}
where $D_{q q} = \lim_{\mathcal{T} \rightarrow \infty} \av{\Delta q_\text{h}^2}/(2\mathcal{T})$ and $D_{w w} = \lim_{\mathcal{T} \rightarrow \infty} \av{\Delta w^2}/(2\mathcal{T})$ characterize the fluctuations of the input heat and output work.
These bounds imply the tradeoff-relation \cite{Pie16,Vro18}
\begin{align}
\av{\dot{w}} \leq \text{min} \bigg( \frac{D_{qq}}{T_\text{c}} \eta \big(\eta_\text{C} - \eta \big), \frac{D_{ww}}{T_\text{c}} \frac{\eta_\text{C} - \eta}{\eta} \bigg) \label{tradeoff-1},
\end{align}
which states that work current has to vanish as the efficiency approaches the Carnot efficiency, assuming the fluctuations of heat and work remain finite.
By contrast, the MTUR \eqref{uncertainty-2dim} yields the bound
\begin{align}
\av{\dot{q}_\text{h}}^2 D_{ww} - &2 \av{\dot{q}_\text{h}} \av{\dot{w}} D_{qw} + \av{\dot{w}}^2 D_{qq} \\
& \leq \frac{1}{T_\text{c}}\big(D_{qq} D_{ww} - D_{q w}^2\big)  (\eta_\text{C} - \eta) \av{\dot{q}_\text{h}} \n .
\end{align}
Replacing $\av{\dot{q}_\text{h}} = \av{\dot{w}}/\eta$ this can be written as
\begin{align}
\av{\dot{w}} \leq \frac{1}{T_\text{c}} \frac{D_{qq} D_{ww} - D_{q w}^2}{\eta^2 D_{qq} - 2 \eta D_{qw} + D_{ww}} \eta (\eta_\text{C} - \eta) \label{tradeoff-2} .
\end{align}
This bound is tighter than the both bounds in Eq.~\eqref{tradeoff-1} for any $\bm{D}$ and $\bm{\eta}$, which are included as limiting cases for $D_{ww} \gg D_{qq}$ and $D_{ww} \ll D_{qq}$, respectively.
Further, it reveals that the correlations $D_{q w} = \lim_{\mathcal{T} \rightarrow \infty} \av{\Delta q_\text{h} \Delta w}/(2\mathcal{T})$ between heat and work play an important role in determining the maximal output power of the engine.
At first sight, it seems that, if input heat and work are strongly correlated, $|D_{qw}|/\sqrt{D_{qq} D_{ww}} \approx 1$, the output power of the engine has to vanish.
This seems contradictory, since, if the operation of the engine is ergodic in the sense that the time-averaged work and heat rates reproduce the ensemble averaged ones, we would expect a one-to-one correspondence between the fluctuations of heat and work in the long-time limit: positive fluctuations of the output work should be compensated by positive fluctuations of the input heat. We thus expect $\av{\Delta q_\text{h} \Delta w}/\sqrt{\av{\Delta q_\text{h}^2} \av{\Delta w^2}} = D_{qw}/\sqrt{D_{qq} D_{ww}} \rightarrow 1$ for long times.
So, in light of the bound Eq.~\eqref{tradeoff-2}, how can any ergodic engine have a finite power output?
As it turns out, the right-hand side vanishes in the limit of strong correlations for any value of $\eta$ except for $\eta = D_{qw}/D_{qq} \simeq \sqrt{D_{ww}/D_{qq}}$.
Thus, for any ergodic engine with finite power output, we necessarily have $\eta = \av{\dot{w}}/\av{\dot{q}_\text{h}} =  \sqrt{D_{ww}/D_{qq}}$.
This implies that such an engine exhibits a particularly simple scaling, where the typical fluctuations of work scale in the same manner as the average, $\av{\dot{w}} \propto \sqrt{D_{ww}}$, and similar for the heat.
Such a simple scaling is often indicative of Gaussian distributions of the heat and work, and we indeed find that this is true for the example studied below.
\begin{figure}
\begin{center}
\includegraphics[width=.48\textwidth]{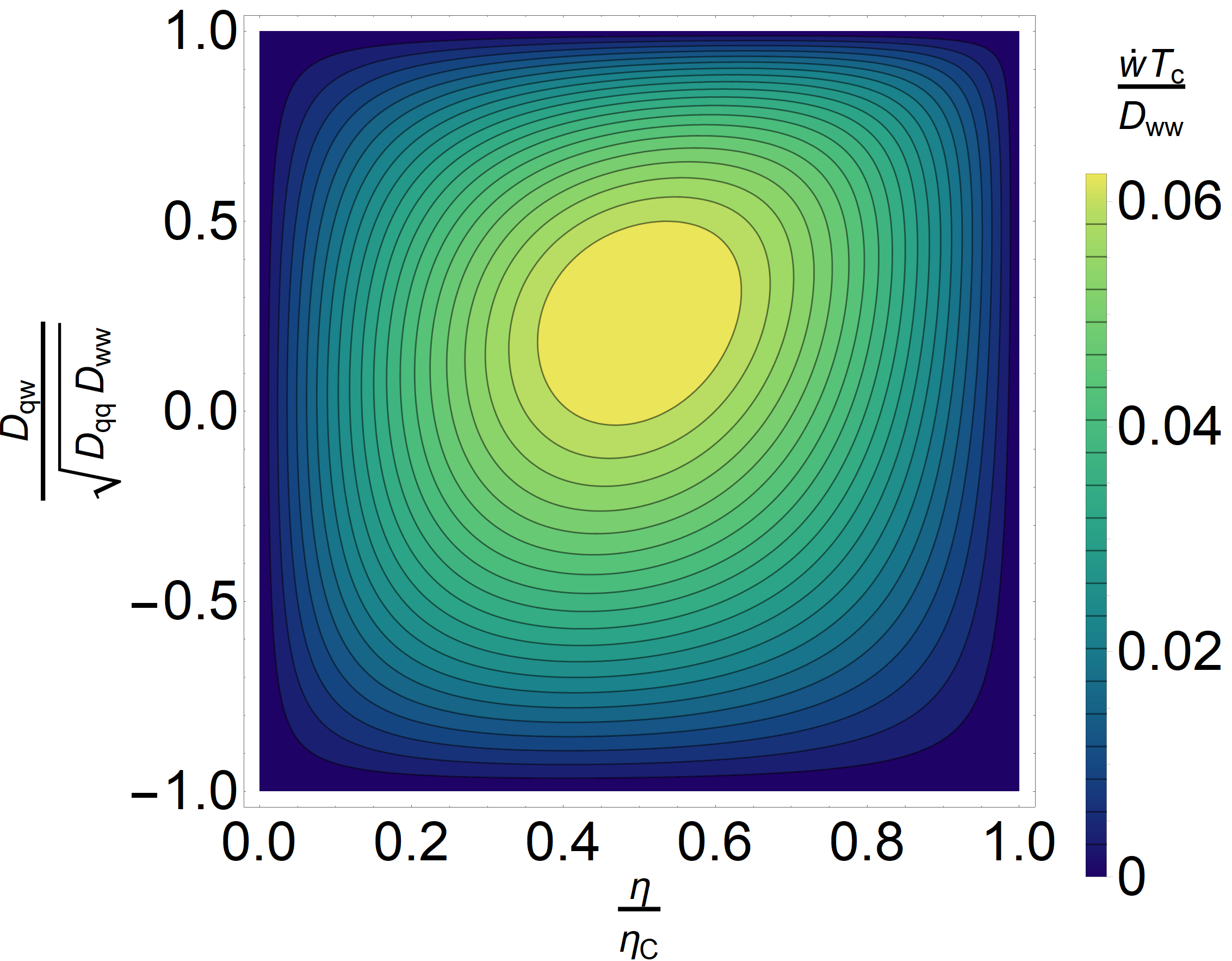}
\includegraphics[width=.48\textwidth]{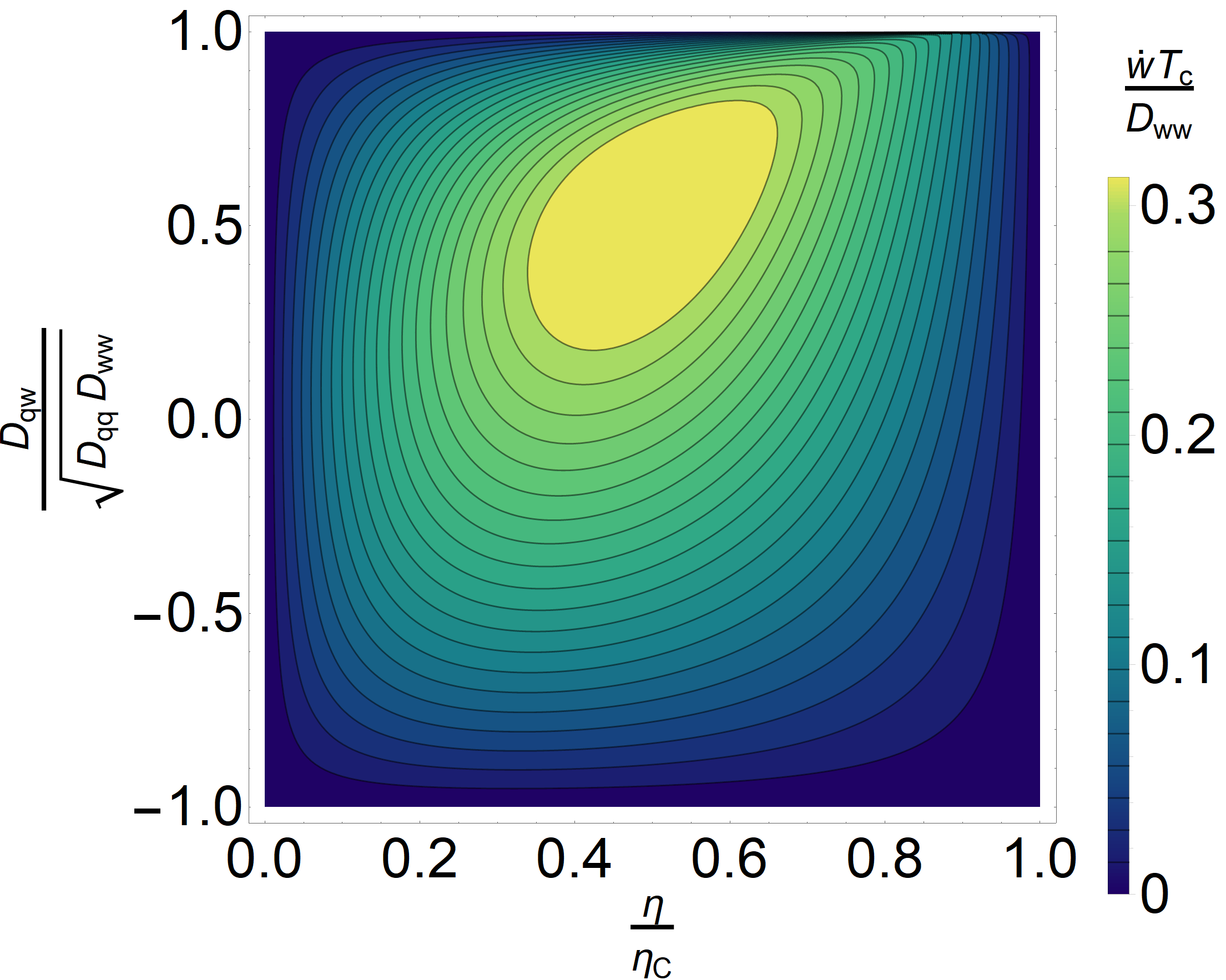}\\
\includegraphics[width=.48\textwidth]{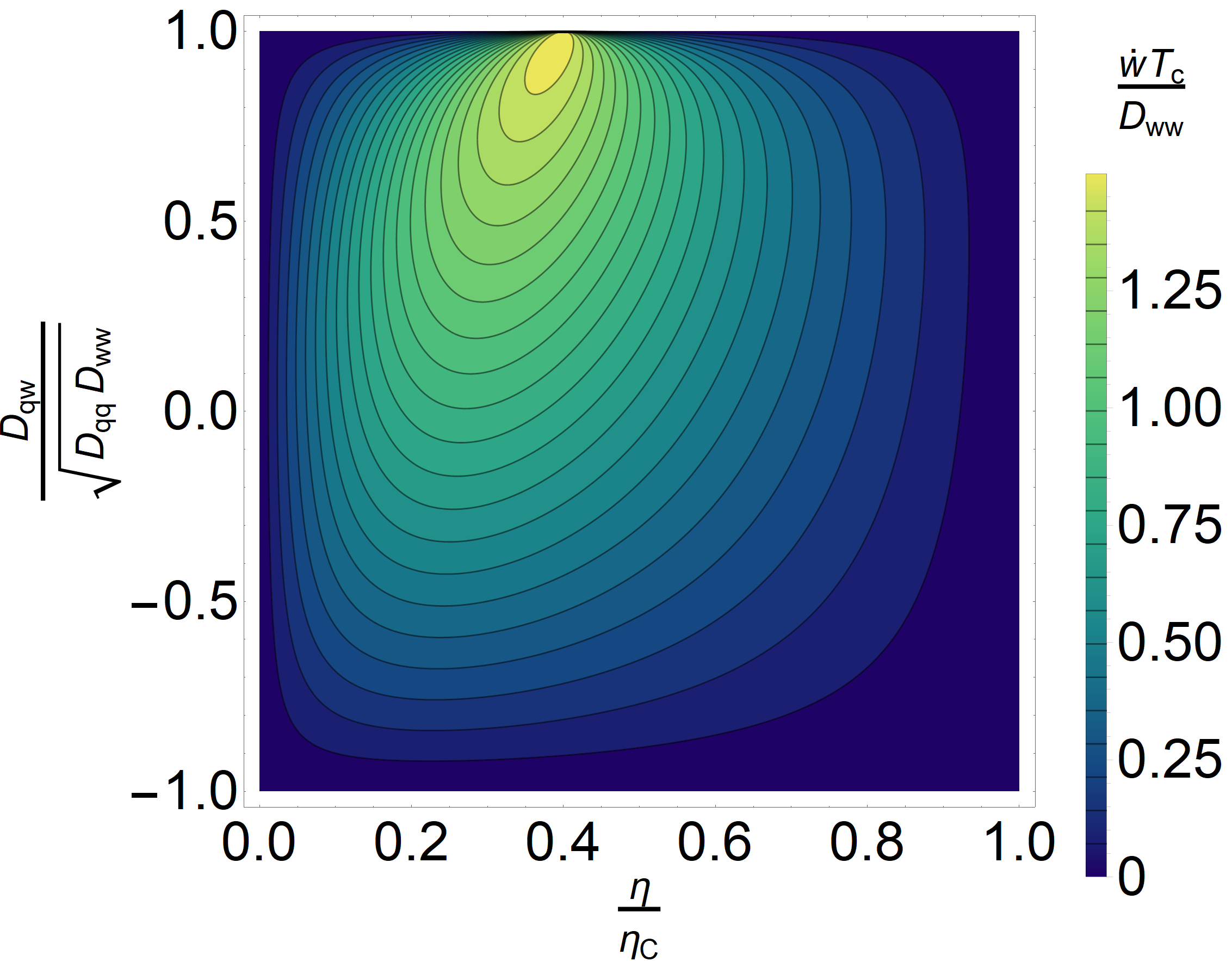}
\end{center}
\caption{The bound Eq.~\eqref{tradeoff-2} on the output power of a stead-state heat engine, as a function of the efficiency relative to the Carnot efficiency $\eta_\text{C} = 0.5$ and the relative size of the correlations between heat and work.
In the top left panel, the fluctuations of heat and work have the same magnitude $D_{qq}/D_{ww} = 1$ and we increase this ratio to $D_{qq}/D_{ww} = 5$ in the top-right panel and $D_{qq}/D_{ww} = 25$ in the bottom panel. The remaining parameters are $D_{ww} = 1$ and $T_\text{c} = 1$. For small to moderate fluctuations of the input heat, the possible power output is small and maximal for weak correlations between heat and work and around $\eta \approx 0.5 \eta_\text{C}$. As the fluctuations of the input heat start to dominate those of the output work, the possible output power increases (note the different scales on the three panels), while the maximum shifts towards stronger correlations and lower efficiencies. Finally, for very large fluctuations of the input heat, the possible output power develops a pronounced maximum for strongly correlated heat and work currents and at $\eta = \sqrt{D_{ww}/D_{qq}}$. \label{fig-tradeoff}}
\end{figure}
In general, we find that, in order to reach a large output power relative to the work fluctuations, the fluctuations in the input heat should dominate those of the work and work and heat should be strongly correlated, see Fig.~\ref{fig-tradeoff}.

To gain some insight into the properties of heat-work correlations, we examine the concrete example of the Landauer-B{\"u}ttiker ratchet \cite{Bue87,Lan88}, which, in its simplest realization, consists of an overdamped particle in a periodic potential and in contact with a heat bath with spatially varying temperature,
\begin{align}
\dot{x}(t) = \mu \big( -U'(x(t)) + F \big) + \sqrt{2 \mu T(x(t))} \cdot \xi(t) \label{bl-ratchet} .
\end{align}
We assume that the potential is periodic with period $\lambda$, $U(x+\lambda) = U(x)$.
As in Ref.~\cite{Dec17}, we take the temperature to vary as $T(x)^{-1} = T_\text{c}^{-1} + (T_\text{h}^{-1} - T_\text{c}^{-1}) \chi(x)$, with a periodic function $\chi(x+\lambda) = \chi(x)$ with $0 \leq \chi(x) \leq 1$.
This system gives rise to a steady state drift velocity, which can be used to perform work against the load force $F$.
In this case, the heat and work can be written as
\begin{align}
q_\text{h}[x(t)] = \int_0^\mathcal{T} dt \ \chi(x(t)) \big( -U'(x(t)) - T'(x(t)) + F \big) \circ \dot{x}(t) \\
w[x(t)] = -F \int_0^\mathcal{T} dt \ \dot{x}(t) \n .
\end{align}
While an analytic computation of the fluctuations of heat and work is challenging, this task is readily accomplished by performing numerical Langevin simulations.
We find that indeed the output work and input heat become strongly correlated in the long-time limit, $\av{\Delta q_\text{h} \Delta w}/\sqrt{\av{\Delta q_\text{h}^2} \av{\Delta w^2}} \rightarrow 1$, while the efficiency takes a value of $\eta \approx \sqrt{D_{ww}/D_{qq}}$, thus avoiding vanishing output power.
Moreover, we find that the output work obeys $\av{\Delta w^4} = 3 \av{\Delta w^2}^2$ for long times, and similar for the input heat, which is indicative of a Gaussian work and heat distribution.
Thus, while the model described by Eq.~\eqref{bl-ratchet} is by no means linear, containing a periodic potential and multiplicative noise, it has a simple, Gaussian and ergodic, structure with respect to heat and work.
Upon closer examination, this is not so surprising, considering that, due to the Markovian nature of the dynamics, the state of the engine is essentially reset whenever the particle travels a distance $L$.
Thus, the contributions to the output work traveling from $0$ to $L$ and from $L$ to $2L$ are independent, and the total work, as a sum of such independent contributions, exhibits a Gaussian distribution in the long-time limit as a consequence of the central limit theorem.

\subsection{TUR and mobility}
In the derivation and discussion of the TUR, we so far used only a single perturbation proportional to the probability current in Eq.~\eqref{cramer-rao}.
However, the general formulation of the Cr{\'a}mer-Rao bound \eqref{cramer-rao} allows us to consider additional perturbations and as a consequence, to connect the TUR to other inequalities.
Let us assume that we are dealing with a steady-state system, so that, similar to the derivation of the TUR, the choice $\bm{a}_1(\bm{x}) = \bm{j}^\text{st}(\bm{x})/P^\text{st}(\bm{x})$ leaves the probability density unchanged.
We further explicitly consider a system of overdamped particles in contact with a heat bath at temperature $T$ with a position-independent (positive definite and symmetric) mobility matrix $\bm{M}$, i.~e.~$\bm{B} = T \bm{M}$,
\begin{align}
\dot{\bm{x}}(t) = \bm{M} \bm{F}(\bm{x}(t)) + \sqrt{2 T \bm{M}} \bm{\xi}(t),
\end{align}
where $\bm{F}(\bm{x})$ is an arbitrary force that may include interactions between the particles and conservative as well as non-conservative external forces.
We now introduce a second perturbation $\bm{a}_2 = \bm{M} \bm{f}$ with a constant force $\bm{f}$ and, as out observable, take the time-integrated displacement vector $\bm{r}[\bm{x}(t)] = \int_0^\mathcal{T} dt \ \dot{\bm{x}}(t)$ with average $\av{\bm{r}}_{\theta} = \mathcal{T} \bm{v}_\theta$, where $\bm{v}_\theta$ is the steady-state drift velocity.
Then, the bound \eqref{cramer-rao}, evaluated at $\theta_1 = \theta_2 = 0$ (i.~e.~in the linear response limit) reads,
\begin{align}
&\left( \begin{array}{c}
\partial_{\theta_1} \bm{v}_{\theta}^T \\[.5 ex]
\partial_{\theta_2} \bm{v}_{\theta}^T
\end{array} \right) \bm{D}^{-1} \left( 
\partial_{\theta_1} \bm{v}_{\theta},
\partial_{\theta_2} \bm{v}_{\theta} \right) \bigg\vert_{\theta = 0} \label{unc-fri-1} \\
&\ \leq \left(\begin{array}{cc}
\sigma^\text{st} & \frac{1}{T}\int d\bm{x} \ \bm{f}^T \bm{M} \bm{M}^{-1} \bm{j}^\text{st}(\bm{x}) \\[.5 ex]
\frac{1}{T}\int d\bm{x} \ \bm{f}^T \bm{M} \bm{M}^{-1} \bm{j}^\text{st}(\bm{x}) & \frac{1}{T} \int d\bm{x} \ \bm{f}^T \bm{M} \bm{M}^{-1} \bm{M} \bm{f} P^\text{st}(\bm{x}) \end{array} \right) \nn
&\ = \left(\begin{array}{cc}
\sigma^\text{st} & \frac{1}{T} \bm{f}^T \bm{v}_0 \\[.5 ex]
\frac{1}{T} \bm{f}^T \bm{v}_0 & \frac{1}{T} \bm{f}^T \bm{M} \bm{f} \end{array} \right) \n ,
\end{align}
where $(\bm{D})_{i j} = \lim_{\mathcal{T} \rightarrow \infty} \av{\Delta r_i \Delta r_j}/(2\mathcal{T})$ is the matrix of diffusivities.
We also define the differential mobility matrix $\bm{\mathcal{M}}$ via
\begin{align}
\partial_{\theta_2} \bm{v}_{\theta} \Big\vert_{\theta = 0} = \bm{\mathcal{M}} \bm{f} .
\end{align}
The elements $(\mathcal{M})_{i j}$ of this matrix encode how much the current in direction $x_i$ changes by applying a small constant force in direction $j$.
Since the perturbation corresponding to $\theta_1$, proportional to the probability current, rescales all currents and thus also the drift velocity to $\bm{v}_\theta \simeq (1+\theta_1) \bm{v}_0$, we have $\partial_{\theta_1} \bm{v}_{\theta} \vert_{\theta=0} = \bm{v}_0$.
The bound Eq.~\eqref{unc-fri-1} is then equivalent to the set of inequalities
\begin{align}
\bm{v}_0^T \bm{D}^{-1} \bm{v}_0 \leq \sigma^\text{st} \label{uncertainty-steady-2} \\
\bm{\mathcal{M}}^T \bm{D}^{-1} \bm{\mathcal{M}} \leq \frac{\bm{M}}{T} \label{mobility-bound} \\
\Bigg(\bm{f}^T \Bigg(\frac{\bm{v}_0}{T} - \bm{\mathcal{M}}^T \bm{D}^{-1} \bm{v}_0 \Bigg) \Bigg)^2 \label{unc-fri-3} \\
\qquad \leq \Big( \sigma^\text{st} - \bm{v}_0^T \bm{D}^{-1} \bm{v}_0 \Big) \bm{f}^T \Bigg( \frac{\bm{M}}{T} - \bm{\mathcal{M}}^T \bm{D}^{-1} \bm{\mathcal{M}} \Bigg) \bm{f} . \n
\end{align}
The first inequality \eqref{uncertainty-steady-2} is exactly the long-time version of the steady-state TUR Eq.~\eqref{uncertainty-steady}.
The second inequality \eqref{mobility-bound} is the bound on the mobility matrix derived in Ref.~\cite{Dec18B}.
The differential mobility $\bm{\mathcal{M}}$ describes the response of the particles to a small force $\bm{f}$ in the presence of the force $\bm{F}$, whereas the bare mobility $\bm{M}$ describes both response and diffusivity in the absence of $\bm{F}$, i.~e.~for free diffusion.
Compared to free diffusion, the force $\bm{F}$ can decrease or increase both the mobility and the diffusivity compared to their bare values.
The inequality \eqref{mobility-bound} states that these changes are not independent of each other but have to follow certain rules:
Increased mobility always is accompanied by enhanced diffusivity; whereas decreased diffusivity can only be achieved at the cost of reduced mobility \cite{Dec18B}.
We remark that this relation was first observed for a particle in a one-dimensional periodic potential and conjectured to hold in more general cases in Ref.~\cite{Hay05}.
We stress that this is valid for arbitrarily strong forces $\bm{F}$ and arbitrarily far from equilibrium.

Finally, the third inequality \eqref{unc-fri-3} shows that the uncertainty relation and the bound on mobility are not independent of each other.
In particular, we can use it to find a condition for a system to saturate the TUR, i.~e.~to have equality in Eq.~\eqref{uncertainty-steady-2}.
This necessarily requires that the left-hand side of Eq.~\eqref{unc-fri-3} vanishes for arbitrary $\bm{f}$, and thus
\begin{align}
T \bm{\mathcal{M}}^T \bm{D}^{-1} \bm{v}_0 = \bm{v}_0 .
\end{align}
This is generically satisfied only if $T \bm{\mathcal{M}}^T \bm{D}^{-1}$ is the identity matrix, i.~e.~for
\begin{align}
\bm{D} = T \bm{\mathcal{M}} ,
\end{align}
which is precisely the equilibrium fluctuation-dissipation relation.
We thus arrive at the following statement:
The TUR for the drift velocity $\bm{v}$ can be an equality only in systems that satisfy the equilibrium fluctuation-dissipation relation.
For a typical non-equilibrium situation, in which the equilibrium fluctuation-dissipation relation is violated, the TUR thus presents a strict inequality.

\section{Discussion}

The multidimensional generalization of the GTUR and TUR developed here allows to investigate the properties of stochastic currents in realistic, high-dimensional transport situations and the interrelations between different stochastic current observables in a natural manner.
On the application side, we have used the MTUR to establish new tradeoff relations for the performance of steady-state heat engines, involving the heat-work correlations.
Such tradeoff relations in various flavors have recently been an active topic in the discussion of stochastic heat engines \cite{Shi16,Pie16,Dec17,Dec18,Pie18,Vro18} and the connection between work fluctuations and output power is essential for understanding if, and under what conditions, finite power at Carnot efficiency is realizable \cite{Cam16,Hol17,Hol18}.
The MTUR shows that not just the fluctuations of work and heat, but also how they are correlated has a strong impact on the output power of an engine.

The derivation of the MTUR from the Cr{\'a}mer-Rao bound reveals that the family of uncertainty relations is actually a consequence of information-theoretic bounds \cite{Has18,Dec18B}.
No observable can contain more information than the underlying probability distribution and thus the effect of a parameter change on an observable is bounded by the Fisher information.
Choosing a suitable perturbation that turns the Fisher information into a physical observable (in case of the TUR the entropy production), the information-theoretic inequality translates into a relation between different physical observables \cite{Dec18B}.

We anticipate that generalizations of other uncertainty relations \cite{Pro17,Chi18,Bra18} to vector-valued observables can be derived in a similar manner.
An open issue, on which some progress has been made recently \cite{Van18}, is the extension of TURs to non-Markovian systems.
We speculate that the understanding of the TUR in terms of information theoretic bounds may provide a useful guideline to derive TURs also in non-Markovian systems.

\section*{Acknowledgments}
This work was supported by the World Premier International Research Center Initiative (WPI), MEXT, Japan.
The author wishes to thank S.-i.~Sasa and S.~Ito for stimulating discussions.

\section*{Appendix: Markov jump processes}
We consider a Markov jump process on a finite state space of $N$ states. 
In this case, the time-evolution of the occupation probabilities $p_k(t)$ with $k = 1, \ldots, N$ is governed by the Master equation
\begin{align}
\partial_t p_k(t) = \sum_{k} \Big( W_{kl}(t) p_l(t) - W_{lk}(t) p_k(t) \Big) \label{master},
\end{align}
where $W_{kl}(t) \geq 0$ are the transition rates from state $l$ to state $k$, which we assume to satisfy the local detailed balance condition $W_{kl}(t) = 0 \Leftrightarrow W_{lk}(t) = 0$.
We take the rates to be governed by a set of parameters
\begin{align}
W_{k l}(t) = W_{k l}^0(t) \exp\Big[ \sum_{\mu = 1}^M \theta_\mu \Omega^\mu_{k l}(t) \Big].
\end{align}
We choose this exponential form with $W_{kl}^0(t) \geq 0$ to ensure that positive rates remain positive for all values of the parameters $\bm{\theta}$.
For short times $\tau$, the transition probability from state $l$ to state $k$ is to leading order given by
\begin{align}
p(k, t+\tau \vert l, t) = \delta_{k l} + \tau \Big( W_{kl}(t) - \delta_{kl} \sum_{m} W_{lm}(t) \Big) + O(\tau^2) .
\end{align}
We can use this to define a path probability on the discretized time interval $[0,\mathcal{T}] = \cup_{n = 1}^N [n \tau, (n-1)\tau]$ with $N\tau = \mathcal{T}$
\begin{align}
\mathbb{P} = \prod_{n=1}^N p(k_n, t_n \vert k_{n-1} t_{n-1}) p_{k_0}(0) .
\end{align}
whose derivative follows by applying the product rule
\begin{align}
\partial_{\theta_\mu} \mathbb{P} = \Bigg( \sum_{n=1}^N \frac{\partial_{\theta_\mu} p(k_n, t_n \vert k_{n-1}, t_{n-1})}{p(k_n, t_n \vert k_{n-1}, t_{n-1})} + \frac{\partial_{\theta_\mu} p_{k_0}(0)}{p_{k_0}(0)} \Bigg) \mathbb{P} .
\end{align}
The Fisher information of the path probability is then given by
\begin{align}
\big(\bm{I}_\theta\big)_{\mu \nu} &= \sum_{k_N} \sum_{k_{N-1}} \ldots \sum_{k_0} \frac{\partial_{\theta_\mu} \mathbb{P} \partial_{\theta_\nu} \mathbb{P}}{\mathbb{P}} \\
&= \sum_{k_N} \sum_{k_{N-1}} \ldots \sum_{k_0} \Bigg( \sum_{n=1}^N \frac{\partial_{\theta_\mu} p(k_n, t_n \vert k_{n-1}, t_{n-1})}{p(k_n, t_n \vert k_{n-1}, t_{n-1})} + \frac{\partial_{\theta_\mu} p_{k_0}(0)}{p_{k_0}(0)} \Bigg) \nn
& \qquad \times \Bigg( \sum_{m=1}^N \frac{\partial_{\theta_\nu} p(k_m, t_m \vert k_{m-1}, t_{m-1})}{p(k_m, t_m \vert k_{m-1}, t_{m-1})} + \frac{\partial_{\theta_\nu} p_{k_0}(0)}{p_{k_0}(0)} \Bigg) \mathbb{P} \n .
\end{align}
We now show that only the diagonal terms in the double sum over $m$ and $n$ contribute.
Consider
\begin{align}
&\sum_{k_N} \sum_{k_{N-1}} \ldots \sum_{k_0} \frac{\partial_{\theta_\mu} p(k_n, t_n \vert k_{n-1}, t_{n-1})}{p(k_n, t_n \vert k_{n-1}, t_{n-1})} \frac{\partial_{\theta_\nu} p(k_m, t_m \vert k_{m-1}, t_{m-1})}{p(k_m, t_m \vert k_{m-1}, t_{m-1})} \mathbb{P} \\
&= \sum_{k_n} \sum_{k_{n-1}} \ldots \sum_{k_0} \partial_{\theta_\mu} p(k_n, t_n \vert k_{n-1}, t_{n-1}) \frac{\partial_{\theta_\nu} p(k_m, t_m \vert k_{m-1}, t_{m-1})}{p(k_m, t_m \vert k_{m-1}, t_{m-1})} \mathbb{P}^{n-1} \n,
\end{align}
where we assumed, without loss of generality, $n \geq m$ and $\mathbb{P}^{n-1}$ is the path probability up to step $n-1$.
This expression is zero for $n > m$ since we have
\begin{align}
\sum_{k_n} \partial_{\theta_\mu} p(k_n, t_n \vert k_{n-1}, t_{n-1}) = \partial_{\theta_\mu} \sum_{k_n}  p(k_n, t_n \vert k_{n-1}, t_{n-1}) = \partial_{\theta_\mu} 1 = 0 
\end{align}
and the only contribution comes from $n = m$, where the expression simplifies to
\begin{align}
&\sum_{k_N} \sum_{k_{N-1}} \ldots \sum_{k_0} \frac{\partial_{\theta_\mu} p(k_n, t_n \vert k_{n-1}, t_{n-1}) \partial_{\theta_\nu} p(k_n, t_n \vert k_{n-1}, t_{n-1})}{p(k_n, t_n \vert k_{n-1}, t_{n-1})^2} \mathbb{P} \\
&= \sum_{k_n} \sum_{k_{n-1}} \frac{\partial_{\theta_\mu} p(k_n, t_n \vert k_{n-1}, t_{n-1}) \partial_{\theta_\nu} p(k_n, t_n \vert k_{n-1}, t_{n-1})}{p(k_n, t_n \vert k_{n-1}, t_{n-1})} p_{k_{n-1}}(t_{n-1}) \n .
\end{align}
The Fisher information is thus additive in the individual steps and can be written as
\begin{align}
\big(\bm{I}_\theta\big)_{\mu \nu} &= \sum_{n=1}^N \Bigg( \sum_{k_n} \sum_{k_{n-1}} \frac{\partial_{\theta_\mu} p(k_n, t_n \vert k_{n-1}, t_{n-1}) \partial_{\theta_\nu} p(k_n, t_n \vert k_{n-1}, t_{n-1})}{p(k_n, t_n \vert k_{n-1}, t_{n-1})} \\
&\hspace{3 cm} \times p_{k_{n-1}}(t_{n-1}) \Bigg)  + \sum_{k_0} \frac{\big(\partial_{\theta_\mu} p_{k_0}(0) \partial_{\theta_\nu} p_{k_0}(0) \big)}{p_{k_0}(0)} \n  .
\end{align}
The derivative of the transition probability with respect to a parameter $\theta_\mu$ is given by
\begin{align}
\partial_{\theta_\mu} p(k, t+\tau \vert l, t) &= \tau \Big( \Omega^\mu_{k l}(t) W_{k l}(t) - \delta_{k l} \sum_m \Omega^\mu_{l m}(t) W_{l m}(t) \Big) \\
& = \tau \frac{Z^\mu_{k l}(t) W_{k l}(t) - \delta_{k l} \sum_m \Omega^\mu_{l m}(t) W_{l m}(t)}{\delta_{k l} + \tau \big( W_{kl}(t) - \delta_{kl} \sum_{m} W_{lm}(t) \big)} p(k, t+\tau \vert l, t) \n .
\end{align}
We can use this to compute the Fisher information for a single step $n - 1 \rightarrow n$,
\begin{align}
\big(\bm{I}_\theta\big)^{n - 1 \rightarrow n}_{\mu \nu} &= \tau^2 \sum_{k_{n}} \sum_{k_{n-1}} \Big( \Omega^\mu_{k_{n} k_{n-1}} W_{k_{n} k_{n-1}} - \delta_{k_{n} k_{n-1}} \sum_m \Omega^\mu_{k_{n-1} m} W_{k_{n-1} m} \big) \Big) \\
&\quad \times \Big( \Omega^\nu_{k_{n} k_{n-1}} W_{k_{n} k_{n-1}} - \delta_{k_{n} k_{n-1}} \sum_m \Omega^\nu_{k_{n-1} m} W_{k_{n-1} m} \Big) \nn
& \quad \times  \frac{p_{k_{n-1}}}{\delta_{k_{n} k_{n-1}} + \tau \big( W_{k_{n} k_{n-1}} - \delta_{k_{n} k_{n-1} \sum_m W_{k_{n} m}}\big)} \n ,
\end{align}
where all time-dependent quantities are evaluated at $t_{n-1}$.
Now, we distinguish the case $k_{n-1} = k_{n}$ and $k_{n-1} \neq k_{n}$, writing
\begin{align}
\big(\bm{I}_\theta\big)^{n - 1 \rightarrow n}_{\mu \nu} &= \tau \sum_{k_{n} \neq k_{n-1}} \sum_{k_{n-1}}  \Omega^\mu_{k_{n} k_{n-1}} \Omega^\nu_{k_{n} k_{n-1}} W_{k_{n} k_{n-1}} p_{k_{n-1}} \\
&\quad + \tau^2 \sum_{k_n} \Big( \Omega^\mu_{k_{n} k_{n}} W_{k_{n} k_{n}} - \sum_m \Omega^\mu_{k_{n} m} W_{k_{n} m} \Big) \nn
&\hspace{2cm}\times \Big( \Omega^\nu_{k_{n} k_{n}} W_{k_{n} k_{n}} - \sum_m \Omega^\nu_{k_{n} m} W_{k_{n} m} \Big) \nn
&\hspace{2cm}\times \frac{p_{k_{n-1}}}{1 + \tau \big( W_{k_{n} k_{n-1}} - \sum_m W_{k_{n} m} \big)} \nn
&= \tau \sum_{k_{n} \neq k_{n-1}} \sum_{k_{n-1}}  \Omega^\mu_{k_{n} k_{n-1}} \Omega^\nu_{k_{n} k_{n-1}} W_{k_{n}, k_{n-1}} p_{k_{n-1}} + O(\tau^2) .
\end{align}
We can then write the Fisher information along the path as
\begin{align}
\big(\bm{I}_\theta\big)_{\mu \nu} &= \tau \sum_{n=1}^N \Bigg( \sum_{k_{n} \neq k_{n-1}} \sum_{k_{n-1}}  \Omega^\mu_{k_{n} k_{n-1}} \Omega^\nu_{k_{n}, k_{n-1}} W_{k_{n} k_{n-1}} p_{k_{n-1}} \Bigg) \\
&\hspace{2cm} + \sum_{k_0} \frac{\big(\partial_{\theta_\mu} p_{k_0}(0) \partial_{\theta_\nu} p_{k_0}(0) \big)}{p_{k_0}(0)} . \n
\end{align}
Taking the continuous-time limit $\tau \rightarrow 0$ with $N \tau = \mathcal{T}$ fixed, the sum can be written as an integral, and we finally obtain,
\begin{align}
\big(\bm{I}_\theta\big)_{\mu \nu} &= \int_0^\mathcal{T} dt \ \sum_{k \neq l}  \Omega^\mu_{k l}(t) \Omega^\nu_{k l}(t) W_{k l}(t) p_{l}(t) + \sum_{k} \frac{\big(\partial_{\theta_\mu} p_{k}(0) \partial_{\theta_\nu} p_{k}(0) \big)}{p_{k}(0)} \label{fisher-path-jump}.
\end{align}
This is the path Fisher information corresponding to the expression Eq.~\eqref{fisher-path} for Langevin dynamics.

Next, we define a stochastic current as
\begin{align}
\bm{r}[k(t)] = \int_0^\mathcal{T} \sum_{k,l} \Big(dn_{k l}(t) - dn_{l k}(t) \Big) \bm{z}_{k l}(t) ,
\end{align}
where $dn_{kl}(t)$ denotes the number of jumps from state $l$ to state $k$ during the time interval $[t,t+\tau]$ and $\bm{z}_{k l}(t)$ is an arbitrary $K$-vector, where $K$ is the number of observables.
The average of such a current is given by
\begin{align}
\av{\bm{r}}_\theta = \int_0^\mathcal{T} dt \ \sum_{k,l} \Big( W_{k l}(t) p_l(t) - W_{l k}(t) p_k(t) \Big) \bm{z}_{k l}(t) .
\end{align}
In order to derive the GTUR, we again need a proper choice of the perturbation $\Omega^1_{k l}$.
Following Ref.~\cite{Koy18}, we take
\begin{align}
\Omega^1_{k l} = \frac{W_{k l}(t) \pi_l - W_{l k}(t) \pi_k}{W_{k l}(t) p^0_l(t) + W_{l k}(t) p^0_k(t)},
\end{align}
where $\pi_k > 0$ is an arbitrary set of positive parameters with $\sum_k \pi_k = 1$ and we denote by $p_k^0(t)$ the solution of the master equation \eqref{master} for $\theta = 0$.
To linear order in $\theta$, the master equation for the modified occupation probabilities is
\begin{align}
\partial_t p_k(t) = &\sum_l \Big( W_{k l}(t) p_l(t) - W_{l k}(t) p_k(t) \Big) \\
& + \theta \Bigg(\frac{W_{k l}(t) \pi_l - W_{l k}(t) \pi_k}{W_{k l}(t) p^0_l(t) + W_{l k}(t) p^0_k(t)} W_{k l}(t) p_l(t) \nn
& \qquad + \frac{W_{k l}(t) \pi_l - W_{l k}(t) \pi_k}{W_{k l}(t) p^0_l(t) + W_{l k}(t) p^0_k(t)} W_{l k}(t) p_k(t) \Bigg) + O(\theta^2) \n .
\end{align}
It is easily verified by direct computation that this is solved by
\begin{align}
p_k(t) = p_k^0(t) + \theta \big( p_k^0(t) - \pi_k \big)
\end{align}
and that the average of the current changes to
\begin{align}
\av{\bm{r}}_\theta = (1+\theta) \av{\bm{r}}_0 .
\end{align}
On the other hand, the Fisher information corresponding to this transformation is
\begin{align}
I(0) = \frac{1}{2} \int_0^\mathcal{T} dt \ \sum_{k \neq l} \frac{\big(W_{k l}(t) \pi_l - W_{l k}(t) \pi_k \big)^2}{W_{k l}(t) p^0_l(t) + W_{l k}(t) p^0_k(t)} .
\end{align}
Using the inequality
\begin{align}
\frac{(a-b)^2}{a+b} \leq \frac{1}{2} (a-b) \ln \bigg(\frac{a}{b}\bigg),
\end{align}
which holds for arbitrary positive numbers $a$ and $b$, we can bound this from above by
\begin{align}
I(0) \leq \frac{1}{2} \int_0^\mathcal{T} dt \ \sum_{k \neq l} \Bigg( \frac{W_{k l}(t) \pi_l - W_{l k}(t) \pi_k }{W_{k l}(t) p^0_l(t) - W_{l k}(t) p^0_k(t)} \Bigg)^2 \sigma_{k l}(t),
\end{align}
with the contribution to the entropy production from the transition from $l$ to $k$,
\begin{align}
\sigma_{k l}(t) = \frac{1}{2} \big( W_{k l}(t) p_l(t) - W_{l k}(t) p_k(t) \big) \ln \Bigg(\frac{W_{k l}(t) p_l(t)}{W_{l k}(t) p_k(t)} \Bigg) .
\end{align}
The upper bound on $I(0)$ is precisely what is termed effective entropy production in Ref.~\cite{Koy18}.
Using the Cr{\'a}mer-Rao bound \eqref{cramer-rao}, we thus obtain the multidimensional GTUR for jump processes,
\begin{align}
\av{\bm{r}}^T \bm{\Xi}_r^{-1} \av{\bm{r}} \leq \frac{1}{2} \Sigma \\
\text{with} \quad \Sigma = \int_0^\mathcal{T} dt \ \sum_{k \neq l} \Bigg( \frac{W_{k l}(t) \pi_l - W_{l k}(t) \pi_k }{W_{k l}(t) p^0_l(t) - W_{l k}(t) p^0_k(t)} \Bigg)^2 \sigma_{k l}(t) . \n
\end{align}
For a steady state process, we can choose $\pi_k = p_k^\text{st}$ and $\Sigma$ reduces to the entropy production $\Delta S = \int_0^\mathcal{T} dt \sum_{k l} \sigma_{k l}(t)$, recovering the MTUR for jump processes,
\begin{align}
\av{\bm{r}}^T \bm{\Xi}_r^{-1} \av{\bm{r}} \leq \frac{1}{2} \Delta S .
\end{align}

\section*{Bibliography}

\bibliographystyle{iopart-num}
\bibliography{bib}

\providecommand{\newblock}{}
\begin{thebibliography}{10}
\expandafter\ifx\csname url\endcsname\relax
  \def\url#1{{\tt #1}}\fi
\expandafter\ifx\csname urlprefix\endcsname\relax\def\urlprefix{URL }\fi
\providecommand{\eprint}[2][]{\url{#2}}

\bibitem{Bar15}
Barato A~C and Seifert U 2015 {\em Phys. Rev. Lett.\/} {\bf 114} 158101
  \urlprefix\url{https://journals.aps.org/prl/abstract/10.1103/PhysRevLett.114.158101}

\bibitem{Gin16}
Gingrich T~R, Horowitz J~M, Perunov N and England J~L 2016 {\em Phys. Rev.
  Lett.\/} {\bf 116}(12) 120601
  \urlprefix\url{https://link.aps.org/doi/10.1103/PhysRevLett.116.120601}

\bibitem{Pie17}
Pietzonka P, Ritort F and Seifert U 2017 {\em Phys. Rev. E\/} {\bf 96}(1)
  012101 \urlprefix\url{https://link.aps.org/doi/10.1103/PhysRevE.96.012101}

\bibitem{Hor17}
Horowitz J~M and Gingrich T~R 2017 {\em Phys. Rev. E\/} {\bf 96}(2) 020103
  \urlprefix\url{https://link.aps.org/doi/10.1103/PhysRevE.96.020103}

\bibitem{Dec17}
Dechant A and i~Sasa S 2018 {\em J. Stat. Mech. Theory E.\/} {\bf 2018} 063209
  \urlprefix\url{http://stacks.iop.org/1742-5468/2018/i=6/a=063209}

\bibitem{Bar18}
Barato A~C, Chetrite R, Faggionato A and Gabrielli D 2018 {\em ArXiv
  e-prints\/} (\textit{Preprint} \eprint{1806.07837})
  \urlprefix\url{https://arxiv.org/abs/1806.07837}

\bibitem{Koy18}
Koyuk T, Seifert U and Pietzonka P 2018 {\em ArXiv e-prints\/}
  (\textit{Preprint} \eprint{1809.02113})
  \urlprefix\url{https://arxiv.org/abs/1809.02113}

\bibitem{Rao45}
Radhakrishna~Rao C 1945 {\em Bull. Calcutta Math. Soc.\/} {\bf 37} 81--91 ISSN
  0008-0659

\bibitem{Cra16}
Cram{\'e}r H 2016 {\em Mathematical methods of statistics\/} vol~9 (Princeton
  university press)

\bibitem{Has18}
{Hasegawa} Y and {Van Vu} T 2018 {\em ArXiv e-prints\/} (\textit{Preprint}
  \eprint{1809.03292}) \urlprefix\url{https://arxiv.org/abs/1809.03292}

\bibitem{Kay93}
Kay S~M 1993 {\em {Fundamentals of statistical signal processing, volume I:
  estimation theory}\/} (Prentice Hall)

\bibitem{Ris86}
Risken H 1986 {\em The Fokker-Planck Equation\/} (Springer Berlin)

\bibitem{Dec18B}
Dechant A and Sasa S~i 2018 {\em ArXiv e-prints\/} (\textit{Preprint}
  \eprint{1804.08250}) \urlprefix\url{https://arxiv.org/abs/1804.08250}

\bibitem{Pie16}
Pietzonka P, Barato A~C and Seifert U 2016 {\em J. Stat. Mech. Theory E.\/}
  {\bf 2016} 124004
  \urlprefix\url{http://stacks.iop.org/1742-5468/2016/i=12/a=124004}

\bibitem{Pie18}
Pietzonka P and Seifert U 2018 {\em Phys. Rev. Lett.\/} {\bf 120}(19) 190602
  \urlprefix\url{https://link.aps.org/doi/10.1103/PhysRevLett.120.190602}

\bibitem{Vro18}
Vroylandt H, Lacoste D and Verley G 2018 {\em J. Stat. Mech. Theory E.\/} {\bf
  2018} 023205
  \urlprefix\url{http://stacks.iop.org/1742-5468/2018/i=2/a=023205}

\bibitem{Bue87}
B{\"u}ttiker M 1987 {\em Z. Phys. B\/} {\bf 68} 161
  \urlprefix\url{https://doi.org/10.1007/BF01304221}

\bibitem{Lan88}
Landauer P 1988 {\em J. Stat. Phys.\/} {\bf 53} 233
  \urlprefix\url{https://doi.org/10.1007/BF01011555}

\bibitem{Hay05}
Hayashi K and Sasa S~i 2005 {\em Phys. Rev. E\/} {\bf 71}(2) 020102
  \urlprefix\url{https://link.aps.org/doi/10.1103/PhysRevE.71.020102}

\bibitem{Shi16}
Shiraishi N, Saito K and Tasaki H 2016 {\em Phys. Rev. Lett.\/} {\bf 117}(19)
  190601
  \urlprefix\url{https://link.aps.org/doi/10.1103/PhysRevLett.117.190601}

\bibitem{Dec18}
Dechant A and Sasa S~i 2018 {\em Phys. Rev. E\/} {\bf 97}(6) 062101
  \urlprefix\url{https://link.aps.org/doi/10.1103/PhysRevE.97.062101}

\bibitem{Cam16}
Campisi M and Fazio R 2016 {\em Nature comm.\/} {\bf 7} 11895

\bibitem{Hol17}
Holubec V and Ryabov A 2017 {\em Phys. Rev. E\/} {\bf 96}(6) 062107
  \urlprefix\url{https://link.aps.org/doi/10.1103/PhysRevE.96.062107}

\bibitem{Hol18}
Holubec V and Ryabov A 2018 {\em Phys. Rev. Lett.\/} {\bf 121}(12) 120601
  \urlprefix\url{https://link.aps.org/doi/10.1103/PhysRevLett.121.120601}

\bibitem{Pro17}
Proesmans K and den Broeck C~V 2017 {\em EPL (Europhysics Letters)\/} {\bf 119}
  20001 \urlprefix\url{http://stacks.iop.org/0295-5075/119/i=2/a=20001}

\bibitem{Chi18}
Chiuchi\`u D and Pigolotti S 2018 {\em Phys. Rev. E\/} {\bf 97}(3) 032109
  \urlprefix\url{https://link.aps.org/doi/10.1103/PhysRevE.97.032109}

\bibitem{Bra18}
Brandner K, Hanazato T and Saito K 2018 {\em Phys. Rev. Lett.\/} {\bf 120}(9)
  090601
  \urlprefix\url{https://link.aps.org/doi/10.1103/PhysRevLett.120.090601}

\bibitem{Van18}
{Van Vu} T and {Hasegawa} Y 2018 {\em ArXiv e-prints\/} (\textit{Preprint}
  \eprint{1809.06610}) \urlprefix\url{https://arxiv.org/abs/1809.06610}

\end{thebibliography}

\end{document}